\documentclass{llncs}

\usepackage{tcolorbox}
\usepackage[color]{coqdoc}
\usepackage{amsmath}

\begin{document}

\title{Interpreted Formalisms for Configurations}

\author{Chong Tang\inst{1} \and
Kevin~Sullivan\inst{1} \and
Jian Xiang\inst{2} \and 
Trent Weiss\inst{1} \and 
Baishakhi Ray\inst{1}}

\authorrunning{Chong Tang et al.} 
\institute{University of Virginia, Charlottesville VA 22904, USA, \\
\email{sullivan@virginia.edu},\\
\and Harvard University, Cambridge, MA, 02138, USA}

\maketitle

\begin{abstract}
Imprecise and incomplete specification of system \textit{configurations} threatens safety, security, functionality, and other critical system properties and uselessly enlarges the configuration spaces to be searched by configuration engineers and auto-tuners. To address these problems, this paper introduces \textit{interpreted formalisms based on real-world types for configurations}.  Configuration values are lifted to values of real-world types, which we formalize as \textit{subset types} in Coq. Values of these types are dependent pairs whose components are values of underlying Coq types and proofs of additional properties about them. Real-world types both extend and further constrain \textit{machine-level} configurations, enabling richer, proof-based checking of their consistency with real-world constraints.   
Tactic-based proof scripts are written once to automate the construction of proofs, if proofs exist, for configuration fields and whole configurations. \textit{Failures to prove} reveal real-world type errors. Evaluation is based on a case study of combinatorial optimization of Hadoop performance by meta-heuristic search over Hadoop configurations spaces. 
\end{abstract}


\section{Introduction}

Configurations are critical elements of many modern software-intensive systems, from big data computing stacks to robots to the \textit{internet of things}. Configurations are collections of parameter values that can be set by end-users to specialize and optimize system functions, performance, and other properties for particular uses or environments. Configurability enables the production of commodity software and software-intensive systems that can be used for diverse purposes. 

Selecting configurations is a fraught exercise. Even individual components can have hundreds of configuration parameters. Systems of systems can have orders of magnitude more.  Configurations are also often under-specified, as manifested in the use of loose \textit{machine-level} types (e.g., \textit{integer, string}), for configuration parameters (or \textit{fields}), and in the incomplete and imprecise specification of constraints on and across fields. These issues often make it unclear what values parameters can reasonably have, what they mean precisely, how to set them to obtain desired system properties (e.g., performance, security), and how not to set them to avoid comprising system properties. 

The complexity, inadequate specification, and opaque meanings of configurations risks the use of bad configurations and vastly enlarges the configuration spaces that configuration engineers and auto-tuners must explore. We propose to address such problems with \textit{interpreted formalisms for configurations}. 

Earlier work by Xiang, Knight and Sullivan \cite{xiang2016synthesis,xiang2017my} identified a lack of explicit, checkable interpretations for code as posing risks to cyber-physical system dependability. They proposed \textit{interpreted formalisms} as a solution. An interpreted formalism augments code with an explicit structure---an \textit{interpretation}, mapped to the code---that imposes \textit{real-world types} on, and further explicates the intended meanings of, code elements, both to aid human understanding and to enable automated checking of code for consistency with real world constraints. 

An interpreted formalism is a $(code, interpretation)$ pair. It can be used to check that machine-level values can be lifted to values of real-world types. Such types can \textit{extend} and further \textit{constrain} machine-type values (e.g., with units, limiting \textit{integer} values to \textit{positive} values, possibly with with additional range restrictions, etc.). Xiang et al.~\cite{xiang2017my} demonstrated the efficacy of interpreted formalisms for finding bugs in Java programs for cyber-physical systems.

The problems we have identified with \textit{configurations} are analogous to those with code. We introduce \textit{interpreted formalisms for configurations} as a solution. We augment  parameters and whole configurations with interpretations to explicate intended meanings and enable checking of configurations against real-world constraints.  We specify real-world types as what amount to dependent pairs in Coq. Values of real-world types combine machine-level values lifted to values of Coq types, with proofs of additionally specified properties of these lifted values. Real-world type checking involves lifting followed by automated construction of proofs. Real-world type errors are detected if either lifted values or constructed proof objects fail to type check in Coq.



As evidence of the feasibility, utility, and conceptual clarity afforded by our approach, we present an interpretation for Apache Hadoop~\cite{apache_hadoop} configurations, including real-world types based on constraints mined from Hadoop documentation. 
The main contributions of this paper can be summarized as follows:
\begin{itemize}
    \item We show that formally specified, fully automated, efficient real-world type checking can be provided for system configurations
    \item We show that real-world type checking can find previously unrecognized errors in  Hadoop configurations
    \item We show that filtering malformed configurations can significantly improvement search efficiency
    \item We show that Coq's dependent type theory and module system support clear, practical, and flexible specification of interpreted formalisms for configurations
    \item We establish foundations for real-world type systems grounded in type theory
\end{itemize}



\section{Background}


In recent work~\cite{xiang2016synthesis,xiang2017my}, Xiang, Knight, and Sullivan identified two major shortcomings in today's software practice. First, software engineers tend to represent properties of real-world phenomena as values of---and in procedures that operate on values of---\textit{under-constrained machine types}. As one example, an altitude relative to ground in meters might be represented only by a value of the machine type, \textit{integer}, perhaps with a name such as \textit{alt} and a comment, \textit{altitude in meters relative to the ground}. The formal type is under-specified in that it permits values, such as $-1$, that are meaningless in the real world. 

The second, closely related, problem is that the intended \textit{interpretations} of code are not specified in a form that enables sufficient automated checking of consistency of code with the real world. Machine-level values and operations are permitted that have no real-world meaning. There is usually nothing to prevent a program from adding an integer (in meters) to an integer (in feet), for example. Similar issues involve frames of reference, staleness of sensor data, measurement error, possibilities for erroneous data from failed sensors, etc.

In order to address these problems, Xiang et al. proposed the concept of the interpreted formalism based on real-world types. In contrast to the current practice, the \textit{real-world type} assigned to {\em alt} might be \textit{non-negative real integer expressed in meters above ground level (AGL)}. The real-world types constrains the value and adds units and a frame of reference. Real-world type systems limit \textit{machine} values to values that are meaningful in the real world while extending them with information critical to the full specification and automated checking of their intended interpretations. In addition to real-world types, an interpretation can include information such as references to relevant standards, expository prose, etc., to further clarify the intended meanings of machine-typed values. 


The present work emerged from an effort in combinatorial optimization of Hadoop performance through novel meta-heuristic  searches for high-performing configurations. We found that the \textit{machine types} of Hadoop configuration parameters (e.g., integer, string, float), and thus of configurations, were often under-constrained, that their intended interpretations were often unclear, and that Hadoop was without mechanisms for checking the values of parameters with real-world constraints. Many fields are documented as being of type integer, for example, even in cases where not any integer will do. We also found some Hadoop documentation to be erroneous. Hadoop's Wiki page\footnote{\url{https://wiki.apache.org/hadoop/HowManyMapsAndReduces}} cites \textit{io.buffer.size} as a configuration field name, but there is no such field. It appears that \textit{io.file.buffer.size} was meant. Among other harms, under-specification enlarges search spaces to include configurations that violate known but unchecked real-world constraints.

\section{Approach}

To address the problems that flow from under-constrained configurations with poorly specified interpretations, we introduce \textit{interpreted formalisms based on real-world types for configurations}. We first describe how we formalize real-world types and lift machine-typed field and configuration values to real-world type checked values. Then we present an example using this mechanism to produce an interpretation for and to type check a Hadoop configuration. 

\subsection{Extending Configurations with Real-World Types} \label{formalrwt}

Configurations, which are collections of constant definitions, are simpler than imperative code. There are usually no assignments to mutable memory, function calls, pointers, sub-typing, etc. Their simplicity has enabled us to clarify our understanding of interpreted formalisms based on real-world types. We formalize a real-world type as a dependent pair type, $(b_r, p_r)$, where $b_r$ is what we have called a \textit{base type} (such as \textit{positive} in Coq), and where $p_r$ is an additional property of values of this type---in Coq, a function from values to propositions about them---such as the property of being divisible by the hardware page size on a given machine. 

Binding a real-world type to a parameter, $p$, with a machine value $v_m$ (such as 65536) of machine type $t_m$ (such as integer), involves the \textit{lifting} of $v_m$ to a corresponding {\em putative} (not yet fully checked) real-world value, $v_r$ (such as 65536\%positive), of type $b_r$ (here $positive$), followed by the construction, \textit{if possible}, of a proof, $c_r$, that this particular putative real-world value, $v_r$, has the additional property $p_r$ (e.g., that \textit{65536\%positive mod 4096\%Z = 0\%Z)}. If a proof, $c_r$, can be constructed, then the dependent pair, $(v_r,c_r)$ can be constructed, and the real-world type of the {\em machine} value, $v_m$, is thereby proved. 

The lift-and-prove operation is essentially a partial function. A machine value $v_m$ \textit{real-world type checks} when it has an image under this function. In further detail, this function takes a given machine term, $(v_m: t_m)$---read as \textit{machine value $v_m$ of machine type $t_m$}---to a real-world term, $(v_r: b_r, c_r: p_r v_r)$---read as \textit{the dependent pair comprising real-world value $v_r$ of (Coq) base type, $b_r$, along with proof, $c_r$, of the proposition, $(p_r~v_r)$, that certifies that $v_r$ has property $p_r$}. Here $c_r$ is a proof term (a value) for the proposition (a type) about $v_r$ to which the Coq property $p_r$ (a function) maps $v_r$. The lift-and-prove function is not defined for $v_m$ if either (1) there is no $v_r$ to which $v_m$ can be lifted, or (2) no proof, $c_r$, can be constructed to certify that $v_r$ has the additional property, $p_r$.

The lifting of a machine value to a putative real-world value generally adds information that is known to the engineer but not explicit in the machine value or type. This additional information is vital for real-world type checking. The addition of constraints on permitted machine values is one example. Another would be that lifting adds information about the physical units in which a machine value is expressed, to enable checking of consistent use of units when machine values are combined. Simple machine types are thus generally lifted to more complex ``base'' types in Coq, to provide room for this added information. For example, we lift machine-level strings representing Hadoop JVM options (such as ``-Xms1024m -Xmx4096m'') to values of record types in Coq with fields of Coq type $positive$ for the numerical values of the initial and maximum virtual machine stack sizes, explicit units (e.g., $m$ for megabytes), and a constraint that the initial value not exceed the maximum value. The lifting operation itself can add and check constraints. For example, attempting to lift the machine-level integer value, $-1$, to a value of the Coq base type $positive$ will fail to type check, irrespective of any additional property of the base-type value that would have to be checked had the lifting succeeded.  

\subsection{Working with Hadoop}

An explicit interpretation when paired with a Hadoop machine-level configuration constitutes an interpreted formalism pair. Our interpreted formalisms precisely specify (1) the previously undocumented parameterization of configurations by \textit{external platform characteristics}, such as the number of hardware CPUs, involved in constraints on the values of Hadoop parameters; (2) units for all relevant parameters, establishing a pattern if augmenting machine types with additional information such as units, frames of reference, etc; (3) all constraints ascertained from both official documentation and other trusted sources, expressed using a combination of (a) base types, such as \textit{positive},  that can be more restrictive than the underlying machine types, and (b) pairing of these lifted values with proofs of additional, declaratively specified properties. 

Coq provides very expressive means for documenting properties (constraints), and powerful facilities for automating much (and in our work to date, all) of the verification of values against such constraints. It also provides trustworthy strong and static verification that all constraints are satisfied, via its foundational type checker. As an example, Hadoop informally documents but does not enforce a constraint that a certain field should have a value that is a multiple of the platform-specific hardware page size. Our interpreted formalism quickly reveals violations of this constraint in failures to generate required proofs. Use cases for such work include (1) automated real-world type checking of configurations, (2) using such type checking to reject mechanically generated, inconsistent configurations prior to costly dynamic profiling, (3) providing a formal specification of the constraints to be satisfied by a future, envisioned, constraint-driven generator of candidate configurations, e.g., using a separate SMT solver, (4) supporting the development of a human-facing interface for improved understanding of complex configurations, which will be critical for human-in-the-loop configuration search/tuning, and (5) for generation of good configurations for use in testing, and of counter-examples for use in fuzz testing. We have already developed (1) through (3) in this paper, with (4)  and (5) left for future work. We are also exploring applications of these ideas to configurations for complex, safety- and security-critical systems, including industrial robots.

\section{Coq Implementation} \label{coqdetails}

This section presents the details of our Coq implementation of real-world types and type checker for Hadoop configuration. 

\subsection{Defined Coq Types}
We begin by instantiating a record type whose fields represent \textit{environment} parameters: parameters not defined as part of Hadoop configurations but that are implicated in constraints on configurations values. For example, the number of CPU cores that MapReduce jobs are permitted to use must not exceed the number of CPUs made available to Hadoop by the hardware and surrounding system, an environment parameter. The following code presents the Coq record \textit{type}. The fields reflect all external parameters that we know to be involved in constraints on the subset of performance-related Hadoop parameters that we have modeled. We elide the imports of libraries for the Coq types used in this code. Details can be found in our GitHub repository at \url{https://github.com/ChongTang/SoS\_Coq}.

\begin{tcolorbox}
\begin{coqdoccode}
\coqdockw{Record} \coqdef{env desc.Env}{Env}{\coqdocrecord{Env}} := \coqdef{env desc.mk env}{mk\_env}{\coqdocconstructor{mk\_env}} \{\coqdoceol
\coqdocindent{2.00em}
\coqdef{env desc.env phys CPU cores}{env\_phys\_CPU\_cores}{\coqdocprojection{env\_phys\_CPU\_cores}}: \coqexternalref{positive}{http://coq.inria.fr/distrib/8.6/stdlib/Coq.Numbers.BinNums}{\coqdocinductive{positive}};\coqdoceol
\coqdocindent{2.00em}
\coqdef{env desc.env virt CPU cores}{env\_virt\_CPU\_cores}{\coqdocprojection{env\_virt\_CPU\_cores}}: \coqexternalref{positive}{http://coq.inria.fr/distrib/8.6/stdlib/Coq.Numbers.BinNums}{\coqdocinductive{positive}};\coqdoceol
\coqdocindent{2.00em}
\coqdef{env desc.env phys mem mb}{env\_phys\_mem\_mb}{\coqdocprojection{env\_phys\_mem\_mb}}: \coqexternalref{positive}{http://coq.inria.fr/distrib/8.6/stdlib/Coq.Numbers.BinNums}{\coqdocinductive{positive}};\coqdoceol
\coqdocindent{2.00em}
\coqdef{env desc.env virt mem mb}{env\_virt\_mem\_mb}{\coqdocprojection{env\_virt\_mem\_mb}}: \coqexternalref{positive}{http://coq.inria.fr/distrib/8.6/stdlib/Coq.Numbers.BinNums}{\coqdocinductive{positive}};\coqdoceol
\coqdocindent{2.00em}
\coqdef{env desc.env hw page size}{env\_hw\_page\_size}{\coqdocprojection{env\_hw\_page\_size}}: \coqexternalref{positive}{http://coq.inria.fr/distrib/8.6/stdlib/Coq.Numbers.BinNums}{\coqdocinductive{positive}};\coqdoceol
\coqdocindent{2.00em}
\coqdef{env desc.env max file desc}{env\_max\_file\_desc}{\coqdocprojection{env\_max\_file\_desc}}: \coqexternalref{positive}{http://coq.inria.fr/distrib/8.6/stdlib/Coq.Numbers.BinNums}{\coqdocinductive{positive}};\coqdoceol
\coqdocindent{2.00em}
\coqdef{env desc.env max threads}{env\_max\_threads}{\coqdocprojection{env\_max\_threads}}: \coqexternalref{positive}{http://coq.inria.fr/distrib/8.6/stdlib/Coq.Numbers.BinNums}{\coqdocinductive{positive}};\coqdoceol
\coqdocindent{2.00em}
\coqdef{env desc.env comp codecs}{env\_comp\_codecs}{\coqdocprojection{env\_comp\_codecs}}: \coqexternalref{list}{http://coq.inria.fr/distrib/8.6/stdlib/Coq.Init.Datatypes}{\coqdocinductive{list}} \coqexternalref{string}{http://coq.inria.fr/distrib/8.6/stdlib/Coq.Strings.String}{\coqdocinductive{string}} \}.\coqdoceol
\end{coqdoccode}
\end{tcolorbox}

We instantiate a record of this type to specify a particular operating environment. In the following code, for example, the list of class names for codecs available in the Java search path on the given platform is encoded as a list of strings. This will enable us later to define and enforce a constraint that a string-valued Hadoop parameter listing codec class names include only values in this list. This environment description record is visible in the parts of our code where one defines constraints on Hadoop field values and whole configurations.

\begin{tcolorbox}
\begin{coqdoccode}
\coqdockw{Definition} \coqdef{env desc.myEnv}{myEnv}{\coqdocdefinition{myEnv}}:\coqref{env desc.Env}{\coqdocrecord{Env}} := \coqref{env desc.mk env}{\coqdocconstructor{mk\_env}} \coqdoceol
\coqdocindent{2.00em}
14\%\coqdocvar{positive} \coqdoceol
\coqdocindent{2.00em}
28\%\coqdocvar{positive}\coqdoceol
\coqdocindent{2.00em}
32768\%\coqdocvar{positive} \coqdoceol
\coqdocindent{2.00em}
32768\%\coqdocvar{positive} \coqdoceol
\coqdocindent{2.00em}
4096\%\coqdocvar{positive}  \coqdoceol
\coqdocindent{2.00em}
3000\%\coqdocvar{positive}  \coqdoceol
\coqdocindent{2.00em}
500\%\coqdocvar{positive}  \coqdoceol
\coqdocindent{2.00em}
("org.apache.hadoop.io.compress.DefaultCodec"::$\ldots$::\coqdocvar{nil}).\coqdoceol
\end{coqdoccode}
\end{tcolorbox}

Next, we formalize real-world types in Coq. As we stated in section \ref{formalrwt}, a real-world type is essentially a dependent pair type, combining a value and a proof of a property about it. We define a type, $RTipe$, the values of which designate the Coq base types for real-world types. These base types are the types to which we will attempt to lift values of concrete machine types extracted from Hadoop configuration files and objects. The mapping from these $RTipe$ values to actual Coq types is given by a function, $typeOfTipe$, elided here. This mechanism allows us to write code that makes decisions based on real-world types, as one cannot \textit{match} on actual types in Coq. Arbitrarily complex Coq types can be used as base types. We use Coq-library-provided string, integer (Z), positive integer (positive), non-negative integer (N), floating point (float), and Boolean (bool) types, along with a record type that we defined to represent values of Java VM options, and an \textit{option positive} type for fields that require either a positive integer value or a special integer, typically $-1$ or $0$, to indicate that an exceptional behavior is required. We could, if necessary, use records that also encode units, frames of reference, and other information critical to explicating and checking real-world types.

\begin{tcolorbox}
\begin{coqdoccode}
\coqdockw{Inductive} 
\coqdef{fieldType.RTipe}{RTipe}{\coqdocinductive{RTipe}} := \coqdef{fieldType.rTipe Z}{rTipe\_Z}{\coqdocconstructor{rTipe\_Z}} \ensuremath{|} \coqdef{fieldType.rTipe pos}{rTipe\_pos}{\coqdocconstructor{rTipe\_pos}} \ensuremath{|} \coqdef{fieldType.rTipe N}{rTipe\_N}{\coqdocconstructor{rTipe\_N}} \ensuremath{|} \coqdef{fieldType.rTipe string}{rTipe\_string}{\coqdocconstructor{rTipe\_string}} \ensuremath{|}
\coqdoceol
\coqdocindent{2.00em}
\coqdef{fieldType.rTipe bool}{rTipe\_bool}{\coqdocconstructor{rTipe\_bool}} \ensuremath{|} \coqdef{fieldType.rTipe JavaOpts}{rTipe\_JavaOpts}{\coqdocconstructor{rTipe\_JavaOpts}} \ensuremath{|} \coqdef{fieldType.rTipe float}{rTipe\_float}{\coqdocconstructor{rTipe\_float}} \ensuremath{|} \coqdef{fieldType.rTipe option pos}{rTipe\_option_positive}{\coqdocconstructor{rTipe\_option\_pos}}.\coqdoceol
\coqdocemptyline
\end{coqdoccode}
\end{tcolorbox}

The core of our design is the parameterized type, $Field$, an instance of which is used to represent a \textit{certified} Hadoop field holding a lifted value for which a requisite proof of the associated property has been provided. The default property imposes no additional constraints. The $Field$ type has two parameters. The first specifies the $RTipe$ of the base type to which a machine value for this field will be lifted. The second specifies the additional property that must hold for any provided value of that base type. A property is represented in Coq as a function from a value of such a  type to a proposition about that value. A $Field$ type thus amounts to a dependent pair type with a few extra fields: (1) \textit{field\_id}: the string name of the Hadoop field (such as \textit{``io.file.buffer.size''}); (2) \textit{field\_final}: a Boolean value indicating whether the field is \textit{final} in the sense of Hadoop, i.e., that the value can't be overridden; (3) \textit{field\_value}: a value of the Coq base type specified by the $RTipe$; and (4) \textit{field\_proof}: a proof that that particular value satisfies the additionally specified property. 

\begin{tcolorbox}
\begin{coqdoccode}
\coqdocnoindent
\coqdockw{Inductive} \coqdef{fieldType.Field}{Field}{\coqdocrecord{Field}} (\coqdocvar{tipe}: \coqref{fieldType.RTipe}{\coqdocinductive{RTipe}}) (\coqdocvar{property}: \coqexternalref{:type scope:x '->' x}{http://coq.inria.fr/distrib/8.6/stdlib/Coq.Init.Logic}{\coqdocnotation{(}}\coqref{fieldType.typeOfTipe}{\coqdocdefinition{typeOfTipe}} \coqdocvariable{tipe}\coqexternalref{:type scope:x '->' x}{http://coq.inria.fr/distrib/8.6/stdlib/Coq.Init.Logic}{\coqdocnotation{)}} \coqexternalref{:type scope:x '->' x}{http://coq.inria.fr/distrib/8.6/stdlib/Coq.Init.Logic}{\coqdocnotation{\ensuremath{\rightarrow}}} \coqdockw{Prop})  := 
\coqdoceol \coqdocindent{1.00em}
\coqdef{fieldType.mk field}{mk\_field}{\coqdocconstructor{mk\_field}} \{\coqdoceol
\coqdocindent{2.00em}
\coqdef{fieldType.field id}{field\_id}{\coqdocprojection{field\_id}}: \coqexternalref{string}{http://coq.inria.fr/distrib/8.6/stdlib/Coq.Strings.String}{\coqdocinductive{string}}; \coqdoceol
\coqdocindent{2.00em}
\coqdef{fieldType.field final}{field\_final}{\coqdocprojection{field\_final}}: \coqexternalref{bool}{http://coq.inria.fr/distrib/8.6/stdlib/Coq.Init.Datatypes}{\coqdocinductive{bool}}; \coqdoceol
\coqdocindent{2.00em}
\coqdef{fieldType.field value}{field\_value}{\coqdocprojection{field\_value}}: (\coqref{fieldType.typeOfTipe}{\coqdocdefinition{typeOfTipe}} \coqdocvariable{tipe}); \coqdoceol
\coqdocindent{2.00em}
\coqdef{fieldType.field proof}{field\_proof}{\coqdocprojection{field\_proof}}: \coqdocvariable{property} \coqref{fieldType.field value}{\coqdocmethod{field\_value}}; \}.\coqdoceol
\coqdocemptyline
\end{coqdoccode}
\end{tcolorbox}

\subsection{Generate Coq Modules from Configuration}

Our next step is to generate one Coq \textit{module} for each Hadoop configuration field to be formalized. Each such module will export the parameterized $Field$ type for the corresponding Hadoop field, a function for creating values of this type, and functions for getting values of the fields of these $Field$ objects, including the Coq base value in a given $Field$ instance. 

We use the Coq module system to generate these modules. To do this, we first define a Coq \textit{module type} (a kind of abstract interface) named \textit{Field\_ModuleType}. The Coq code is elided here. It specifies what field-specific information has to be provided for each field to generate the required module. We then generate one intermediate module, conforming to this interface, for each Hadoop field to be formalized. We automate this process with a Python script. Each such module provides field-specific data: the Hadoop field name (a string), its $RTipe$ and thus indirectly its Coq base type, the additional property that the value of this type must satisfy, measurement units (if any), and two strings, one for a natural language explication of the meaning of the field, and another for guidance on how to set the field value. Our Python script maps machine types to $RTipe$ specifications in each such module, stubbing out the additional properties to be \textit{fun value \ensuremath{\Rightarrow} True} and stubbing out the remaining fields, which we don't yet use, to be empty strings. We hand-edit these modules to specify any more restrictive field-level constraints (e.g., here that the $io.file.buffer.size$ value should be divisible by the hardware page size). Here is an example.

\begin{tcolorbox}
\begin{coqdoccode}
\coqdocnoindent
\coqdockw{Module} \coqdocvar{io\_file\_buffer\_size\_desc} <: \coqdocvar{Field\_ModuleType}.\coqdoceol
\coqdocindent{1.00em}
\coqdockw{Definition} \coqdocvar{fName} := "io.file.buffer.size".\coqdoceol
\coqdocindent{1.00em}
\coqdockw{Definition} \coqdocvar{rTipe} := \coqdocvar{rTipe\_pos}.\coqdoceol
\coqdocindent{1.00em}
\coqdockw{Definition} \coqdocvar{rProperty} := \coqdockw{fun} \coqdocvar{value}: \coqdocvar{positive} \ensuremath{\Rightarrow} \\
\coqdocindent{2.00em} 
((\coqdocvar{Zpos} \coqdocvar{value}) \coqdocvar{mod} (\coqdocvar{Zpos} (\coqdocvar{myEnv}.(\coqdocvar{env\_hw\_page\_size})))) = 0\%\coqdocvar{Z}.\coqdoceol
\coqdocindent{1.00em}
\coqdockw{Definition} \coqdocvar{fUnit} := "".\coqdoceol
\coqdocindent{1.00em}
\coqdockw{Definition} \coqdocvar{fInterp} := "".\coqdoceol
\coqdocindent{1.00em}
\coqdockw{Definition} \coqdocvar{fAdvice} := "".\coqdoceol
\coqdocnoindent
\coqdockw{End} \coqdocvar{io\_file\_buffer\_size\_desc}.\coqdoceol
\end{coqdoccode}
\end{tcolorbox}

Finally, we run each such module through a \textit{module functor} to produce the required module for the given field (details elided). These modules provide the $Field$ types and associated functions used in constructing and accessing values encoded in $Field$ objects. Details can be found in the source code.

Having formalized Hadoop fields, we now formalize the types of \textit{multi-field configurations} as record types with fields whose types are the $Field$ types exported by these per-field modules. The following code, for example, formalizes Hadoop's \textit{core-config} configuration. Each field has the same name as its corresponding Hadoop field except that dots are replaced by underscores due to Coq naming conventions. The type of each field is specified to be the $Field$ type exported by the corresponding field module. A value of this type will then represent an actual, concrete, certified Hadoop \textit{core} configuration object. 

\begin{tcolorbox}
\begin{coqdoccode}
\coqdocnoindent
\coqdockw{Record} \coqdef{core config.CoreConfig}{CoreConfig}{\coqdocrecord{CoreConfig}} := \coqdef{core config.mk core config}{mk\_core\_config}{\coqdocconstructor{mk\_core\_config}} \{\coqdoceol
\coqdocnoindent
\coqdocindent{2em}
\coqdef{core config.io file buffer size}{io\_file\_buffer\_size}{\coqdocprojection{io\_file\_buffer\_size}}: \coqdocdefinition{io\_file\_buffer\_size.ftype};\coqdoceol
\coqdocindent{2em}
\coqdef{core config.io map index interval}{io\_map\_index\_interval}{\coqdocprojection{io\_map\_index\_interval}}: \coqdocdefinition{io\_map\_index\_interval.ftype};\coqdoceol
\coqdocindent{2em}
\coqdef{core config.io map index skip}{io\_map\_index\_skip}{\coqdocprojection{io\_map\_index\_skip}}: \coqdocdefinition{io\_map\_index\_skip.ftype};\coqdoceol
\coqdocindent{2em}
\coqdef{core config.io seqfile compress blocksize}{io\_seqfile\_compress\_blocksize}{\coqdocprojection{io\_seqfile\_compress\_blocksize}}: \coqdocdefinition{io\_seqfile\_compress\_blocksize.ftype};\coqdoceol
\coqdocindent{2em}
\coqdef{core config.io seqfile sorter recordlimit}{io\_seqfile\_sorter\_recordlimit}{\coqdocprojection{io\_seqfile\_sorter\_recordlimit}}: \coqdocdefinition{io\_seqfile\_sorter\_recordlimit.ftype};\coqdoceol
\coqdocindent{2em}
\coqdef{core config.ipc maximum data length}{ipc\_maximum\_data\_length}{\coqdocprojection{ipc\_maximum\_data\_length}}: \coqdocdefinition{ipc\_maximum\_data\_length.ftype}\}.\coqdoceol
\end{coqdoccode}
\end{tcolorbox}

Whereas we specify constraints on individual field values within $Field$ objects, we specify constraints on whole configurations by including in their type definitions extra fields of \textit{propositional types}.  As an example, at the end of MapReduce configuration type we specify a multi-field constraint saying that the maximum size of the input data chunk must be greater than the minimum size. In this way, we have fully formalized the real-world types of configurations for Hadoop's core, HDFS, Yarn, and Map-Reduce components and of overall Hadoop configurations. Here's an example of the kind of constraint we can specify for configuration objects.

\begin{tcolorbox}
\begin{coqdoccode}
\coqdocnoindent
\coqdef{mapred config.maxsplit lt minsplit}{maxsplit\_lt\_minsplit}{\coqdocprojection{maxsplit\_lt\_minsplit}}: \coqdoceol \coqdocindent{2em}\coqexternalref{Z.gt}{http://coq.inria.fr/distrib/8.6/stdlib/Coq.ZArith.BinInt}{\coqdocdefinition{Z.gt}} (\coqexternalref{Zpos}{http://coq.inria.fr/distrib/8.6/stdlib/Coq.Numbers.BinNums}{\coqdocconstructor{Zpos}} (\coqdocdefinition{mapreduce\_input\_fileinputformat\_split\_maxsize.value} \coqdoceol
\coqdocindent{7.0em} \coqref{mapred config.mapreduce input fileinputformat split maxsize}{\coqdocmethod{mapreduce\_input\_fileinputformat\_split\_maxsize}})) \coqdoceol
\coqdocindent{3.5em} (\coqexternalref{Z.of N}{http://coq.inria.fr/distrib/8.6/stdlib/Coq.ZArith.BinInt}{\coqdocdefinition{Z.of\_N}} (\coqdocdefinition{mapreduce\_input\_fileinputformat\_split\_minsize.value} \coqdoceol
\coqdocindent{7.0em} \coqref{mapred config.mapreduce input fileinputformat split minsize}{\coqdocmethod{mapreduce\_input\_fileinputformat\_split\_minsize}}))\coqdoceol
\end{coqdoccode}
\end{tcolorbox}

\subsection{Initialize and check Configuration}

We now use a Python script to lift Hadoop configurations to values of Coq configurations types to type check them. Lifted configurations look much like real configuration files. See the following example, in which we use the \textit{mk\_yarn\_config} constructor to instantiate a Coq configuration object, $a\_yarn\_config$, of type $YarnConfig$. For each field, we generate
a call to the \textit{mk} function from the per-field $Field$ module to instantiate a $Field$ object of the requisite type, providing the required values for its components: (1) a Boolean value specifying whether the value is final or not (the false's); (2) a field value, now of a value of the required Coq base type; and (3) a proof object to prove that the value of the field satisfies the properties specified for that value, but using an underscore as a \textit{hole} for a proof to be constructed using Coq tactics. We provide additional proof objects, again as
\textit{holes}, for the cross-field constraints (elided here). The whole definition is wrapped in a Coq \textit{unshelve refine} tactic, with a tactic-based proof building script at the end that fills in the required proof objects if it's possible to construct them. 

\begin{tcolorbox}
\begin{coqdoccode}
\coqdocnoindent
\coqdockw{Definition} \coqdef{a hadoop config.a yarn config}{a\_yarn\_config}{\coqdocdefinition{a\_yarn\_config}}: \coqdocrecord{YarnConfig}.\coqdoceol
\coqdocnoindent
\coqdockw{Proof}.\coqdoceol
\coqdocnoindent
\coqdocvar{unshelve} \coqdoctac{refine} (\coqdoceol
\coqdocindent{2.00em}
\coqdocconstructor{mk\_yarn\_config}\coqdoceol
\coqdocindent{3.00em}
(\coqdocdefinition{yarn\_nodemanager\_container\_\_manager\_thread\_\_count.mk} 
\coqdoceol
\coqdocindent{6.00em}
\coqexternalref{false}{http://coq.inria.fr/distrib/8.6/stdlib/Coq.Init.Datatypes}{\coqdocconstructor{false}}   20\%\coqdocvar{positive} \coqdocvar{\_} )\coqdoceol

\coqdocindent{3.00em}
$\ldots$\coqdoceol

\coqdocindent{3.00em}
(\coqdocdefinition{yarn\_sharedcache\_admin\_thread\_\_count.mk}  
\coqdoceol
\coqdocindent{6.00em}
\coqexternalref{false}{http://coq.inria.fr/distrib/8.6/stdlib/Coq.Init.Datatypes}{\coqdocconstructor{false}}   1\%\coqdocvar{positive} \coqdocvar{\_} )\coqdoceol

\coqdocindent{3.00em}
$\ldots$
); 
\coqdoceol
\coqdoctac{try} (\coqdoctac{exact} \coqexternalref{I}{http://coq.inria.fr/distrib/8.6/stdlib/Coq.Init.Logic}{\coqdocconstructor{I}}); \coqdoctac{try} \coqdoctac{compute}; \coqdoctac{try} \coqdoctac{reflexivity}; \coqdoctac{auto}.\coqdoceol
\coqdocnoindent
\coqdockw{Qed}.
\end{coqdoccode}
\end{tcolorbox}


We specify a real-world type for an entire Hadoop configuration as a Record whose fields are values of the real-world types of the four Hadoop subsystems. We anticipate that the methods developed here can be adapted to     deeply hierarchically structured configurations for large and complex systems.

\begin{tcolorbox}
\begin{coqdoccode}
\coqdocnoindent
\coqdockw{Record} \coqdef{hadoop config.HadoopConfig}{HadoopConfig}{\coqdocrecord{HadoopConfig}} := \coqdef{hadoop config.mk hadoop config}{mk\_hadoop\_config}{\coqdocconstructor{mk\_hadoop\_config}} \{\coqdoceol
\coqdocindent{2.00em}
\coqdef{hadoop config.yarn config}{yarn\_config}{\coqdocprojection{yarn\_config}}: \coqdocrecord{YarnConfig};\coqdoceol
\coqdocindent{2.0em}
\coqdef{hadoop config.mapred config}{mapred\_config}{\coqdocprojection{mapred\_config}}: \coqdocrecord{MapRedConfig};\coqdoceol
\coqdocindent{2.0em}
\coqdef{hadoop config.core config}{core\_config}{\coqdocprojection{core\_config}}: \coqdocrecord{CoreConfig};\coqdoceol
\coqdocindent{2.0em}
\coqdef{hadoop config.hdfs config}{hdfs\_config}{\coqdocprojection{hdfs\_config}}: \coqdocrecord{HDFSConfig}\}.\coqdoceol
\end{coqdoccode}
\end{tcolorbox}

Given a complete, machine-level Hadoop configuration, with core, map-reduce, Yarn, and HDFS sub-configurations, our Python script lifts it to a corresponding value of this $HadoopConfig$ type. In this way, machine-type field and whole configuration values that encode real-world concepts get converted to values of real-world types that make their full real-world meanings explicit and subject to mechanical checking for real-world consistency.


\section{Evaluation}

We now consider the extent to which this work makes the contributions claimed in the introduction.

\subsection{An Advance in Real-World Type Systems}

This work has demonstrated the feasibility and effectiveness of constructing interpreted formalisms based on real-world types for complex configurations. It has shown how Coq's type system can be used to define real-world types that clearly express the essential properties of otherwise inadequately typed machine values. As an example, Hadoop encodes values of what are essentially \textit{option positive} real-world types as mere \textit{integers}, with either $0$ or $-1$ (inconsistently) representing \textit{None}. Coq's parameterized algebraic data types (such as \textit{option T}), and its \textit{propositions as types} paradigm, enable the highly expressive representation and trustworthy checking of an unlimited range of real-world types. Representing real world types as Coq types rather than as the simple and somewhat inflexible record types in the original work of Xiang et al. represents a significant advance over the prior state of the art in real-world type systems.


\subsection{Detecting Real-World Errors in Configurations}
One of the main purposes of a real-world type system is to reveal inconsistencies in software that elude machine-level type systems. Our case study demonstrates the potential for real-world type systems to find inconsistencies in configurations. The context of this paper is a project on meta-heuristic search through spaces of configurations. Our work to date generates Hadoop
configurations in spaces spanned by the specifications of a few machine-typed values to be considered for each Hadoop parameter. Unfortunately, not every combination of machine-type values make sense in the real world. Interposing our real-world type checker between our  configuration generator and the costly experimental profiling operation allows us to greatly improve search performance by eliminating many configurations from consideration before subjecting them to costly experimental evaluation. Here are a few concrete examples.



As one example, the machine type of \textit{mapreduce.jobtracker.maxtasks.perjob} is integer, where a positive value imposes a resource limit and $-1$ means no limit. Our generator was programmed to allow this field value to vary between $-1$ and $4$ based on the machine type of the field. A problem is that a value of $0$ actually makes no sense for this field, as that would indicate that the maximum number of tasks that can be allocated to a given job is zero. Adding a constraint that the field not be $0$, which we did by lifting the field to the real-world \textit{option positive} type, eliminated many nonsensical configurations from consideration. Lifting $0$ to $Some 0\%positive$ yields a Coq term that simply doesn't type check.  




Using properties to further constraint lifted terms of Coq base types also revealed real-world inconsistencies. The formula $min(min\_splitsize, min(blocksize,\\max\_splitsize)$, for example, is used to compute the chunk size in Hadoop, where $bloacksize$ is the size of a data block in HDFS. If the $min\_splitsize$ is greater than $max\_splitsize$, the final chunk size will be the smaller of the values of $blocksize$ and $max\_splitsize$, which is semantically wrong. Although a MapReduce job won't fail because of this error, it will behave in unexpected ways. Our type checker finds violations of this constraint.

Another cross-field constraint violation that our type checker found to our surprise had to do with a set of four constraints about Hadoop's \textit{uber mode}. The constraints are documented in Hadoop's official documentation~\footnote{\url{https://hadoop.apache.org/docs/r2.7.4/hadoop-mapreduce-client/hadoop-mapreduce-client-core/mapred-default.xml}}. They say that if users enable \textit{uber mode}, the CPU and memory resources of \textit{map} and \textit{reduce} tasks must be less than those of the application master. 

It is not surprising that adding constraints invalidates some, or even many, configurations. The concept of constraint-driven design space exploration isn't new. A more interesting implication is that what we should be doing is to base our configuration generator on the real-world types of configurations rather than on their machine types! Consider again the \textit{mapreduce.jobtracker.maxtasks.perjob} field. A $-1$ value iindicates not just another numerical limit, but rather is a flag indicating "no limit is imposed." A generator should treat "no limit" as fundamentally different than $1$ or $2$ or $3$. A multi-level exploration strategy is then called for---either no limit or one of a range of numerical values. Proper consideration of the \textit{real-world} types of field can inform meta-heuristic search strategies, a point we plan to pursue further in future work.

\subsection{Net Improvement in Meta-Heuristic Search Performance}
\label{hadoop-case-study}
To produce a data point on how filtering constraint-violating configurations can improve search performance, we used our real-world type checker to type-check $5,000$ randomly generated configurations, of the kind we generate and test in our search methods. $1,293$ were invalid. One invocation of our runtime Hadoop performance profiling operation takes about $30$ seconds. We run each job $3$ times to obtain an average performance measurement. The saved time is the difference between the time needed to dynamically evaluate $1,293$ configurations and the time needed to type-check $5,000$ configurations. The time to dynamically profile Hadoop running under $1,293$ configurations was about $1293*30*3 = 116370$ seconds. Each type check takes about $0.63$ seconds. The total time to check $5,000$ configurations was thus about $5000*0.63 = 3150$ seconds. The saved time was $116370-3150 = 113220$ seconds out of a total time of $5000*30*3 = 450000$ seconds. The saved time is about $113220/450000 \approx 0.25$, or $25\%$ of the total search time. Specifying and checking informally and often incompletely documented constraints on configurations can clearly reduce search spaces and improve search efficiency significantly.

\subsection{A Flexible Real-World Type System for Configurations}

Our real-world type system for Hadoop configurations has been easy to use. We wrote a  Python script to (1) instantiate $Field$ meta-data modules for each Hadoop configuration field, based on a spreadsheet, in which for each field we entered information about field name, machine type, Coq base type, and natural language explications of intended interpretations along with guidance for configuration engineers, and (2) generate all associated configuration type specifications. Once this code is synthesized, the remaining tasks are to edit additional properties in-by-hand and to create and check configuration objects, which we do by automatically running the $coqc$ command-line Coq type checker on the generated files. It is easy to add and extend real world types to the system: on the order of an hour of work in our experience.

\subsection{Precise Formal Specification of Configuration Spaces}
Our specification of the real-world type of a Hadoop configuration provides an authoritative formal specification of this configuration space, and as a template for specifications of other such configuration spaces. It precisely specifies of the set of all and only valid Hadoop configurations, limited here to a subset of about 100 performance-related fields. In particular, we formalize configuration spaces as types in the constructive logic of Coq. This work enables a precise specification the optimization problem that motivated this work: find \textit{argmin (c: HadoopConfig) runtime(b,c),} where $c$ encodes a configuration in a particular context, $b$ is a benchmark Hadoop job, and $HadoopConfig$ is the real-world type of Hadoop configurations. Optimizing system quality attributes by searching over dependently typed representations thus emerges as a fundamental mathematical problem formulation that seems worthy of further consideration.

\section{Related Work}

The approach proposed in this paper is related to several research areas. We summarize them in this section. 

\textbf{Interpreted formalisms}. This paper advances the theory of interpreted formalisms and real-world types~\cite{xiang2016interpreted} with a formalization based on type theory. This approach makes the expressiveness of higher-order constructive logic available for defining and checking real-world types. Such a checker can be used to establish comprehensive properties. 

\textbf{Type systems}. Pluggable type system \cite{dietl2011building} provide the capability to impose additional type rules on code. Compared with them, our approach exploits the expressive power of dependent types, here with configurations as the "base code" to be further checked. 

\textbf{Configuration errors}. Finding configuration errors has been an active research topic. Mechanisms can be categorized as reactive or proactive. Reactive mechanisms use postmortem analysis of erroneous behaviors and check configuration settings against predefined constraints. Proactive mechanisms try to automatically predict and stop configuration errors early by using techniques such as emulation~\cite{xu2016early}, inference~\cite{xu2017automatic}, and learning~\cite{santolucito2016probabilistic,yuan2011context,zhang2014encore}. Our pro-active mechanism is unique in exploiting real-world types to exclude configuration errors.

\textbf{Performance optimization}. Optimizing system performance by configuration search and tuning is not a new idea. Duan et al.~\cite{duan2009tuning} proposed to improve database performance by auto-tuning configurations, for example. They sample and profile configurations in a cycle-stealing manner, aborting configuration profiling operations that exceed runtime limits. A type-checker such as ours promises to save significant time in such applications. Configuration search has been used in many domains: energy and delay optimization in embedded hardware~\cite{palermo2005multi}; to reduce cache flushing~\cite{zhang2004self}); robot motion planning~\cite{jaillet2010sampling}; and for connectivity problems~\cite{burns2005toward}. Many approaches account for constraints. Our work is novel in bringing type theory and proof engineering to bear on both expressing and checking constraints. 




\section{Conclusion and Future Work}

This paper provides engineering foundations for configuration specification and certification.  It opens up a range of possibilities for future work.


\textbf{Configuration safety and security.} We aim to extend this work to address the need to configure systems to improve a range of critical properties beyond runtime performance. System safety and security are high on our list of priorities. Security can easily be compromised by deployment of security-suboptimal or just simply broken configurations. We postulate that our approach to scalable and efficient real-world type checking of configurations provides an effective basis for expressing and checking complex constraints that must first be learned and then enforced in given environments to assure that critical systems properties such as security are obtained.

\textbf{Trustworthy reconfiguration}: In many systems, environment parameter values change dynamically, potentially invalidating or de-optimizing given configurations. We plan to explore ongoing real-world checking of evolving configurations. Long-running systems might also balance the exploitation of current best-known configurations with cycle-stealing exploration for better ones, again with strong assurance that only valid configurations will be explored.  

\textbf{Learning constraints}: The ability to evaluate configurations dynamically opens up the possibility of learning constraints at runtime. This fits well with systems for which a high-level goal is known \textit{a priori}, e.g. \textit{the drone should not crash}, but configuration values that achieve this goal are not known, e.g. that \textit{the drone should not fly faster than X meters per second}, for some X. We envision that learned invariants will be added dynamically to real-world type system specifications as a kind of machine learning to perform better.


\textbf{Human-in-the-loop configuration search}: The recent Optometrist algorithm~\cite{baltz2017nuclearfusion} places a human in a meta-heuristic loop searching for better configurations for a nuclear fusion plasma containment system. With our approach, we envision an additional possible role for human experts in the configuration search loop: using manual proof engineering to discharge proof obligations that remain after automated proof finders make as much progress as they can.

\textbf{Dependently typed fields}: One key property of configurations that we did not address in this paper is that the real-world types of some fields can sometimes depend on the real-world values of other fields. As an example, if a particular Boolean-valued parameter value is set to true, indicating that some function is enabled, then an entire sub-configuration might be needed for that function, otherwise the configuration field could be set to $void$. Configurations are dependently typed in this sense. We are actively working to adapt the approach in this paper to support configuration spaces with such features.

\textbf{Formalization for imperative code}: We gained a great deal of insight by forcing ourselves to be formal about the nature of the lifting and checking operations of a real-world type system as described in Section 3.1. Having garnered these insights about interpreted formalisms and real-world types in the simplified domain of configurations, we are eager to determine how to port our insights back to the realm of real-world type systems for imperative code. A critical issue will be to demonstrate consistency of real-world type checking with the semantics of the underlying programming language.


\subsubsection*{Acknowledgements}

The work that led to this paper was supported in part by grants from the U.S. Department of Defense, the Systems Engineering Research Center (a U.S. Department of Defense University Affiliated Ressearch Center), and from the National Science Foundation. This paper is dedicated to the memory of John Knight, who was instrumental in developing and evaluating the concept of interpreted formalisms based on real-world types.

\bibliography{bibliography}{}

\begin{thebibliography}{10}
\providecommand{\url}[1]{\texttt{#1}}
\providecommand{\urlprefix}{URL }

\bibitem{apache_hadoop}
Apache hadoop. \url{http://hadoop.apache.org/} (2017), accessed: 2016-08-06

\bibitem{baltz2017nuclearfusion}
Baltz, E., Trask, E., Binderbauer, M., Dikovsky, M., Gota, H., Mendoza, R.,
  Platt, J., Riley, P.: Achievement of sustained net plasma heating in a fusion
  experiment with the optometrist algorithm. Scientific Reports  7 (2017)

\bibitem{burns2005toward}
Burns, B., Brock, O.: Toward optimal configuration space sampling. In:
  Robotics: Science and Systems. pp. 105--112. Citeseer (2005)

\bibitem{dietl2011building}
Dietl, W., Dietzel, S., Ernst, M.D., Mu{\c{s}}lu, K., Schiller, T.W.: Building
  and using pluggable type-checkers. In: Proceedings of the 33rd International
  Conference on Software Engineering. pp. 681--690. ACM (2011)

\bibitem{duan2009tuning}
Duan, S., Thummala, V., Babu, S.: Tuning database configuration parameters with
  ituned. Proceedings of the VLDB Endowment  2(1),  1246--1257 (2009)

\bibitem{jaillet2010sampling}
Jaillet, L., Cort{\'e}s, J., Sim{\'e}on, T.: Sampling-based path planning on
  configuration-space costmaps. IEEE Transactions on Robotics  26(4),  635--646
  (2010)

\bibitem{palermo2005multi}
Palermo, G., Silvano, C., Zaccaria, V.: Multi-objective design space
  exploration of embedded systems. Journal of Embedded Computing  1(3),
  305--316 (2005)

\bibitem{santolucito2016probabilistic}
Santolucito, M., Zhai, E., Piskac, R.: Probabilistic automated language
  learning for configuration files. In: International Conference on Computer
  Aided Verification. pp. 80--87. Springer (2016)

\bibitem{xiang2016interpreted}
Xiang, J.: Interpreted Formalism: Towards System Assurance and the Real-World
  Semantics of Software. Ph.D. thesis, University of Virginia (2016)

\bibitem{xiang2016synthesis}
Xiang, J., Knight, J., Sullivan, K.: Synthesis of logic interpretations. In:
  High Assurance Systems Engineering (HASE), 2016 IEEE 17th International
  Symposium on. pp. 114--121. IEEE (2016)

\bibitem{xiang2017my}
Xiang, J., Knight, J., Sullivan, K.: Is my software consistent with the real
  world? In: High Assurance Systems Engineering (HASE), 2017 IEEE 18th
  International Symposium on. pp. 1--4. IEEE (2017)

\bibitem{xu2016early}
Xu, T., Jin, X., Huang, P., Zhou, Y., Lu, S., Jin, L., Pasupathy, S.: Early
  detection of configuration errors to reduce failure damage. In: OSDI. pp.
  619--634 (2016)

\bibitem{xu2017automatic}
Xu, X., Li, S., Guo, Y., Dong, W., Li, W., Liao, X.: Automatic type inference
  for proactive misconfiguration prevention. In: Proceedings of the 29th
  International Conference on Software Engineering and Knowledge Engineering
  (2017)

\bibitem{yuan2011context}
Yuan, D., Xie, Y., Panigrahy, R., Yang, J., Verbowski, C., Kumar, A.:
  Context-based online configuration-error detection. In: Proceedings of the
  2011 USENIX conference on USENIX annual technical conference. pp. 28--28.
  USENIX Association (2011)

\bibitem{zhang2004self}
Zhang, C., Vahid, F., Lysecky, R.: A self-tuning cache architecture for
  embedded systems. ACM Transactions on Embedded Computing Systems (TECS)
  3(2),  407--425 (2004)

\bibitem{zhang2014encore}
Zhang, J., Renganarayana, L., Zhang, X., Ge, N., Bala, V., Xu, T., Zhou, Y.:
  Encore: Exploiting system environment and correlation information for
  misconfiguration detection. ACM SIGPLAN Notices  49(4),  687--700 (2014)

\end{thebibliography}
\bibliographystyle{splncs03} 

\end{document}